\documentstyle[preprint,prd,aps]{revtex}
\tightenlines
\begin{document}

%
 
\let\a=\alpha      \let\b=\beta       \let\c=\chi        \let\d=\delta   
\let\e=\varepsilon \let\f=\varphi     \let\g=\gamma      \let\h=\eta       
\let\k=\kappa      \let\l=\lambda     \let\m=\mu               
\let\o=\omega      \let\r=\varrho     \let\s=\sigma    
\let\t=\tau        \let\th=\vartheta  \let\y=\upsilon    \let\x=\xi       
\let\z=\zeta       \let\io=\iota      \let\vp=\varpi     \let\ro=\rho        
\let\ph=\phi       \let\ep=\epsilon   \let\te=\theta  
\let\n=\nu  
\let\D=\Delta   \let\F=\Phi    \let\G=\Gamma  \let\L=\Lambda  
\let\O=\Omega   \let\P=\Pi     \let\Ps=\Psi   \let\Si=\Sigma  
\let\Th=\Theta  \let\X=\Xi     \let\Y=\Upsilon  
 
%
 
%
 
\def\cA{{\cal A}}                \def\cB{{\cal B}} 
\def\cC{{\cal C}}                \def\cD{{\cal D}} 
\def\cE{{\cal E}}                \def\cF{{\cal F}} 
\def\cG{{\cal G}}                \def\cH{{\cal H}}                  
\def\cI{{\cal I}}                \def\cJ{{\cal J}}                 
\def\cK{{\cal K}}                \def\cL{{\cal L}}                  
\def\cM{{\cal M}}                \def\cN{{\cal N}}                 
\def\cO{{\cal O}}                \def\cP{{\cal P}}                 
\def\cQ{{\cal Q}}                \def\cR{{\cal R}}                 
\def\cS{{\cal S}}                \def\cT{{\cal T}}                 
\def\cU{{\cal U}}                \def\cV{{\cal V}}                 
\def\cW{{\cal W}}                \def\cX{{\cal X}}                  
\def\cY{{\cal Y}}                \def\cZ{{\cal Z}} 
%
 
\newcommand{\Ns}{N\hspace{-4.7mm}\not\hspace{2.7mm}} 
\newcommand{\qs}{q\hspace{-3.7mm}\not\hspace{3.4mm}} 
\newcommand{\ps}{p\hspace{-3.3mm}\not\hspace{1.2mm}} 
\newcommand{\ks}{k\hspace{-3.3mm}\not\hspace{1.2mm}} 
\newcommand{\des}{\partial\hspace{-4.mm}\not\hspace{2.5mm}} 
\newcommand{\desco}{D\hspace{-4mm}\not\hspace{2mm}}


\draft
\title{Phenomenology of a Quark Mass Matrix from Six Dimensions and its implication for the
Strong CP problem} 
\author{P. Q. Hung\cite{email1}, M. Seco\cite{email2}, A. Soddu\cite{email3}}
\address{Dept. of Physics, University of Virginia, \\
382 McCormick Road, P. O. Box 400714, Charlottesville, Virginia 22904-4714}
\date{\today}
\maketitle

\begin{abstract}

A model of quark mass matrices from six dimensions, which is nearly
democatic in nature and which is previously constructed
by two of us (PQH and MS), is studied in detail in this manuscript. We found
that not only it fits all the six quark masses as well as the CKM
matrix but also that there exists a region in the allowed parameter
space of the model where the constraint on the parameter
$\bar{\theta}$ of the Strong CP problem is satisfied. This region
itself puts a constraint on the CKM parameters $\bar{\rho}$ and
$\bar{\eta}$. As such, through our analysis, there appears to be
a deep connection between Strong and Weak CP in this model.

\end{abstract}


\section{Introduction}

The search for a plausible model of fermion masses is a continuing quest
in particle physics. In particular, the quark sector is a fertile ground
to test various models since it is there that one has the largest
amount of information: quark masses, CKM angles, CP phase, and it is in
this sector that most models turn their attention to.

Two of us have recently constructed a model of fermion masses 
\cite{PQMarcos} in which the mass matrix is almost of the
pure phase form and is constructed out of four plus two
extra compact spatial dimensions. As shown in \cite{PQMarcos},
one extra compact spatial dimension was needed to give
a democratic mass matrix and another one was needed to make its
matrix elements complex. In \cite{PQMarcos}, an almost 
pure phase mass matrix was found to
take the form $M=g_Yv/\sqrt{2}{(1-\r_{ij})exp(i\theta_{ij})}$ with 
$i,j=1,2,3$, $\r_{ij}<<1$, and $\theta_{ij}<<1$, although
Ref. \cite{PQMarcos} contains a more general result.
Our motivation for that work was based on an attempt to
give a theoretical basis for the so-called pure phase mass matrix (PPMM)
ansatz (similar to the previous form but with
$\r_{ij}=0$) of Ref.\cite{branco,Fishbane:xa} which, 
at the time of its construction, was
quite successful in fitting the various angles and masses. In so doing,
we arrived at a mass matrix which contains the pure phase
form as a particular limit. 
As we shall see below, the general result of Ref. \cite{PQMarcos} 
allows us to be able
to fit the latest determination of the CKM elements \cite{Buras}. Along the way,
as stipulated in Ref. \cite{PQMarcos}, we found a special region,
in the allowed parameter space that fits the CKM matrix, where 
the parameter $\bar{\theta}$ of the famous strong CP problem can be
found to satisfy the experimental bound $\bar{\theta} < 10^{-9}$.
This result is somewhat surprising since it is not at all evident
that solutions of our model that fit the mass spectrum and the CKM
matrix could also give values of $\bar{\theta}$ below the
experimental bound. This connection between weak and strong
CP is certainly very intriguing and will be the subject of our focus
at the end of the paper.

The organization of the paper will be as follows. We first briefly
review the construction of a democratic mass matrix (DMM) in 
five dimensions. We then summarize
the salient points of the model of Ref. \cite{PQMarcos}: its 
construction in six dimensions and the resulting quark mass matrices.
Next we enumerate and describe the parameters used in the numerical
analysis of the mass matrices which is carried out in the section that 
follows. There we will show the allowed region in our parameter
space where solutions are found to fit both the mass spectrum
and the CKM matrix. Finally, we discuss a subspace in the allowed
region where the bound $\bar{\theta} < 10^{-9}$ is obeyed. In
particular, we present some thoughts on the possible physics
which might be responsible for this behaviour.

\section{Democratic Mass Matrices from five dimensions}

Before discussing the results of Ref. \cite{PQMarcos}, 
let us first review how a democratic mass matrix (DMM) \cite{dmm}
arises in the case with one extra compact dimension. A DMM
is a special case of an almost PPMM with
$\r_{ij}=0, \,\theta_{ij}=0$, namely one in which all matrix
elements are unity, apart from a common factor, and hence the
name ``democratic''.
As described in \cite{PQMarcos}, in order to obtain a democratic mass matrix and
to avoid unwanted Flavour-Changing Neutral Current processes (FCNC), we
imposed the following permutation symmetries on the Action:
$S_3^Q \otimes S_3^{U^c} \otimes S_3^{D^c}$,
with $Q \rightarrow S_3^Q Q$, 
$U^c \rightarrow S_3^{U^c} U^c$ and $D^c \rightarrow S_3^{D^c} D^c$. 
$Q(x^{\a},y),\,  U^c(x^{\a},y), \, D^c(x^{\a},y)$ 
are the five-dimensional Dirac fields whose left-handed zero modes are given respectively by 
$q(x^{\a})$, $u^c_R(x^{\a})$, and $d^c_R(x^{\a})$. (For convenience, left-handed fields
are used throughout \cite{PQMarcos} and in this paper with the last two fields
representing actually the two quark $SU(2)_L$ singlets.)
The extra dimension is compactified on an $S_1/Z_2$ orbifold.
The action which obeys this permutation symmetry is the sum of two
terms, $S_0$ and $S_{Yuk}$, where

\begin{eqnarray}
S_0& = & \int d^5x \,\bar{Q}_i(i\desco_5 + f\phi(y))Q_i + \bar{U}^c_i(i\desco_5 + f\phi(y)-m_U)U^c_i 
\nonumber \\
& & +\bar{D}^c_i(i\desco_5 + f\phi(y)-m_D)D^c_i \, ,
\label{S0}
\end{eqnarray}
 
\begin{equation}
S_{Yuk} = \int d^5x k_U\sum_iQ_i^TC_5H\sum_j U_j^c + \int d^5x k_D\sum_iQ_i^TC_5\tilde{H}\sum_j D_j^c + h.c. \, .
\label{SYuk}
\end{equation}

In Eq. (\ref{S0}) $D_5$ is the covariant derivative. (The gauge fields are supposed to be uniformly 
spread along the fifth dimension $y$ inside the thick brane.) 
$\phi(y)\delta_{ij}$ 
is the vacuum expectation value VEV for the background scalar field $\Phi(x^{\a},y)$.
The attractive proposal of \cite{ArkaniSchmaltz} to localize chiral zero modes at
different points along the extra dimension $y$ was adopted in \cite{PQMarcos}. As a result,
$m_U$ and $m_D$ are the five dimensional ``masses'' 
which determine the locations of $u^c_R(x^{\a})$ and $d^c_R(x^{\a})$ along $y$.
(As pointed out 
in Ref. \cite{PQMarcos}, in order to have an invariant ``mass term'' under the $Z_2$ symmetry,  
one has to require  
a ``mass reversal'' $m \rightarrow -m$. The behaviour of $m$ under $Z_2$ 
could come for example in a 
model where the ``masses'' are generated by the radiative corrections to the VEV $\phi(y)$, 
$\phi(y) \rightarrow \phi(y) + \d\phi$ with $\d\phi$ being independant of $y$. Because 
$\Phi(x^{\a},y) \rightarrow -\Phi(x^{\a},L_5-y)$ under $Z_2$ symmetry, at the same time 
$\d\phi\rightarrow -\d\phi$ originating in this ``mass reversal''.) 
In Eq. (\ref{SYuk}) $k_U$ and $k_D$ are the Yukawa couplings in five dimensions 
which have been chosen real and flavor 
independant and $H(x^{\a},y)$ is the five dimensional SM doublet Higgs field whose zero mode 
$h(x^{\a})$ is assumed to be uniformly spread along $y$ inside the thick brane. Here 
$C_5=\g_0\g_2\g_y$ is the charge conjugation operator in five dimensions. 

For the purpose of keeping track of the dimensionalities of various objects, let us
remind ourselves that, in five dimensions, the Yukawa coupling $k_U$ has a (mass)
dimension $M^{-1/2}$. A scalar field, in five dimensions, has a dimension $M^{3/2}$.
The zero mode of the SM Higgs field can be written as $K\,h(x^{\a})$ where
$h(x^{\a})$ is the usual 4-dimensional Higgs field with dimension $M$, and therefore
the constant $C$ has a dimension $M^{1/2}$. In dimensionally reducing the above
action to four dimensions, the following dimensionless combination appears 
in the Yukawa term:
\begin{equation}
\label{coupling}
g_{Y,u} \equiv k_{U}\, K\,,
\end{equation}

The effective action for the Yukawa term of the Up sector in four dimensions can now be written as

\begin{equation}
S_{Yuk}^{eff} = \int d^4x g_{Y,u}\sum_{i,j}q_i^T(x^{\a})h(x^{\a})Cu_j^c(x^{\a})\int dy \xi_q^i(y) \xi_{u^c}^j(y) 
+  h.c. \, .
\label{SYukeff}
\end{equation}    

\noindent
where a similar expression holds for the Down sector.
As stressed in Ref. \cite{PQMarcos}, since all the $q_i$'s are located at the same 
place inside the brane, and similarly for all the $u_i^c$'s. The wave function overlap 
$ \int dy \xi_q^i(y) \xi_{u^c}^j(y) $ is independant of $i,j$.
The Yukawa action now becomes 

\begin{equation}
S_{Yuk}^{eff} = \int d^4x g_{Y,u}^{eff} q^T(x^{\a})h(x^{\a}) \left(\begin{array}{ccc}
1 & 1 & 1 \\           
1 & 1 & 1 \\
1 & 1 & 1 
\end{array}\right) Cu^c(x^{\a})  
+  h.c. \, ,
\label{SYukeff2}
\end{equation} 

\noindent
where $g_{Y,u}^{eff}$ is given by 

\begin{equation}
g_{Y,u}^{eff} = g_{Y,u} \int dy \xi_q(y) \xi_{u^c}(y) \, .
\label{gYu}
\end{equation}

\noindent
and similarly for the down sector

\begin{equation}
g_{Y,d}^{eff} = g_{Y,d} \int dy \xi_q(y) \xi_{d^c}(y) \, .
\label{gYd}
\end{equation}

\noindent
From Eq. (\ref{SYukeff2}) one obtains the democratic mass matrix (DMM)  

\begin{equation}
M=g_Y\frac{v}{\sqrt{2}}\left(\begin{array}{ccc}
1 & 1 & 1 \\           
1 & 1 & 1 \\
1 & 1 & 1 
\end{array}\right) \, ,
\label{DMM1}
\end{equation}

\noindent
which has eigenvalues $3g_Yv/\sqrt{2}, 0, 0$, with $v \sim 246 GeV$. 
The DMM of Eq. (\ref{DMM1}) does not reproduce the right 
mass spectrum and the right CKM matrix. What has been shown in 
Ref. \cite{PQMarcos} is that by adding another compact extra 
dimension one can obtain a viable scenario represented by an almost-PPMM.

\section{Almost Pure Phase Mass Matrices from Six Dimensions}   
 
\subsection{The Action}

The main idea of Ref. \cite{PQMarcos} is that by introducing a sixth compact extra dimension, 
and by requiring that the background scalar field couples to the 
fermions through a Yukawa interaction which is non-local along
that extra dimension, one can obtain an oscillatory behavior
for the fermion wave function along the sixth dimension. 
Fermions are delocalized along the sixth dimension, in contrast with the fifth dimension, 
and the oscillatory behavior of the wave functions, together with the breaking of family symmetry, 
has the effect of producing phases in the mass matrix. Let us now
summarize the main results obtained in Ref. \cite{PQMarcos}.
We first rewrite more compactly the action given in Ref. \cite{PQMarcos} which is
the sum of $S_0$ and $S_{Yuk}$ where  

\begin{eqnarray}
S_0& = & \int d^6x \,[\bar{Q}_i i\desco_6 Q_i 
+ \bar{Q}_i(z)(\frac{f}{2}\phi_i(z) - \frac{m_{Q,i}}{2})Q_i(L_6-z) - \bar{Q}_i(L_6-z)
(\frac{f}{2}\phi_i(z) - \frac{m_{Q,i}}{2})Q_i(z) 
\nonumber \\
& &
+ \bar{U}^c_i i\desco_6 U^c_i 
+ \bar{U}^c_i(z)(\frac{f}{2}\phi_i(z) - \frac{m_{U,i}}{2})U^c_i(L_6-z) - \bar{U}^c_i(L_6-z)
(\frac{f}{2}\phi_i(z) - \frac{m_{U,i}}{2})U^c_i(z)
\nonumber \\
& & +\bar{D}^c_i i\desco_6 D^c_i 
+ \bar{D}^c_i(z)(\frac{f}{2}\phi_i(z) - \frac{m_{D,i}}{2})D^c_i(L_6-z) - \bar{D}^c_i(L_6-z)
(\frac{f}{2}\phi_i(z) - \frac{m_{D,i}}{2})D^c_i(z) 
\nonumber \\
& &
+ f^{\prime}\bar{Q}_i\G_7\phi^{\prime}(y)Q_i 
+ \bar{U}^c_i\G_7(f^{\prime}\phi^{\prime}(y) - m_U)U^c_i 
+ \bar{D}^c_i\G_7(f^{\prime}\phi^{\prime}(y) - m_D)D^c_i] \, ,
\label{S0six}
\end{eqnarray}

\noindent
and

\begin{equation}
S_{Yuk} = \int d^6x k_U\sum_iQ_i^TC_6H\sum_j U_j^c + \int d^6x k_D\sum_iQ_i^TC_6\tilde{H}\sum_j D_j^c + h.c. \, ,
\label{SYuksix}
\end{equation}

\noindent
where $C_6= \G_0 \G_2 \G_z$ is the charge conjugation in six dimensions.
(The gamma matrices in six dimensions can be obtained in Ref. \cite{PQMarcos}.) 
In Eq. (\ref{S0six}) we expressed the dependance from $z$ only for the non-local interaction terms. 
The important point here is that while these interactions will produce an oscillatory 
behavior for the fermion wave functions along the sixth dimensions, the local terms, which are built
using $\G_7$, are found to be responsible for localizing the fermions along the fifth dimension. 
The above actions are invariant under the family permutation symmetry
$S_3^Q \otimes S_3^{U^c} \otimes S_3^{D^c}$.

The vacuum expectation values (VEV) for the background scalar fields $\Phi$ and $\Phi^{\prime}$ are
given by

\begin{equation}
<\Phi(x^{\a},y,z)> = \left(\begin{array}{ccc}
\phi_1(z) & 0 & 0 \\           
0 & \phi_2(z) & 0 \\
0 & 0 & \phi_3(z) 
\end{array}\right) \,,
\label{thickz}
\end{equation}
 
\noindent
and

\begin{equation}
<\Phi^{\prime}(x^{\a},y,z)> = \left(\begin{array}{ccc}
\phi^{\prime}(y) & 0 & 0 \\           
0 & \phi^{\prime}(y) & 0 \\
0 & 0 & \phi^{\prime}(y) 
\end{array}\right) \,.
\label{thicky}
\end{equation}
 
As in Ref. \cite{PQMarcos}, the family symmetry is broken by the 
background scalar field $\phi_i(z)\d_{ij}$ and by introducing different non-local ``mass terms'' $m_{Q,i}$, 
$m_{U,i}$ and $m_{D,i}$ for each family. To break the family symmetry together with the left right 
symmetry along the sixth dimension  will allow us to reproduce the right mass spectrum and the
right CKM matrix. 

As shown in Ref. \cite{PQMarcos}, the absence in Eq. (\ref{S0six}) of local interactions of 
the form $\bar{\Psi} \Phi \Psi$, which will localize the fermions wave function along the sixth dimension, 
is obtained by introducing the discrete symmetry Q   

\begin{eqnarray}
\psi(x^{\a},z) &\rightarrow& Q\psi(x^{\a},z) = \G_7 \psi(x^{\a},z) \, ,\nonumber \\
\psi(x^{\a},L_6-z) &\rightarrow & Q\psi(x^{\a},L_6-z) = -\G_7 \psi(x^{\a},L_6-z) \, ,\label{Qsym}\\
\Phi(x^{\a},z) &\rightarrow & Q\Phi(x^{\a},z)= \Phi(x^{\a},z) \, . \nonumber
\end{eqnarray}  

\noindent
As pointed out in Ref. \cite{PQMarcos} the realization of the Q-symmetry of Eq. (\ref{Qsym}) 
implies that the introduced orbifold for the compactification is $S_1/(Z_2 \times Z_2^{\prime})$ 
instead of $S_1/Z_2$.  
This also implies that the physical space is $[0,L_6/2]$ instead of the initial support $[0,L_6]$.  
The non-local terms of Eq. (\ref{S0six}) and the local terms containing $\G_7$ are invariant under the above 
Q-symmetry.

\subsection{The Mass Matrix}

From the Yukawa action of Eq. (\ref{SYuksix}) one can now obtain 
the effective action in four dimensions 

\begin{equation}
S_{Yuk}^{eff} = \int d^4x g_{Y,u}\sum_{i,j}q_i^T(x^{\a})h(x^{\a})Cu_j^c(x^{\a})
\int dy \xi_{5,q}^i(y) \xi_{5,u^c}^j(y) \int dz \xi_{6,q}^{i*}(z) \xi_{6,u^c}^j(z)
+  h.c. \, .
\label{SYukeff6}
\end{equation}  

\noindent
where we considered only the up sector, but equal considerations hold for the down sector. 
Using Eqs. (\ref{SYukeff2}) and (\ref{gYu}) one can rewrite $S_{Yuk}^{eff}$ as 

\begin{equation}
S_{Yuk}^{eff} = \int d^4x g_{Y,u}^{eff} q^T(x^{\a})h(x^{\a}) \left(\begin{array}{ccc}
a_{11} & a_{12} & a_{13} \\           
a_{21} & a_{22} & a_{23} \\
a_{31} & a_{32} & a_{33} 
\end{array}\right) Cu^c(x^{\a})  
+  h.c. \, ,
\label{SYukeff62}
\end{equation}

\noindent
where 

\begin{eqnarray}
a_{ij} & = & \int dz \xi_{6,q}^{i*}(z) \xi_{6,u^c}^j(z) \nonumber \\
& = & \frac{1}{L_6} \int_0^{L_6} dz \,
exp[-i(2fV_iln(cosh(\m_iz))/{\m_i} - 2fV_jln(cosh(\m_jz))/{\m_j} \nonumber \\
& & 
-(m_{q,i} - m_{u,j})z)]
\label{aijused}
\end{eqnarray}

\noindent
In Eq. (\ref{aijused}) we have used for $\xi_{6,q}^i$ and $\xi_{6,u^c}^i$ respectively

\begin{equation}
\xi_{6,q}^i = \frac{1}{\sqrt{L_6}}exp[i(2fV_iln(cosh(\m_iz))/\m_i-m_{q,i}z)]
\label{wavesixq}
\end{equation}

\begin{equation}
\xi_{6,u^c}^i = \frac{1}{\sqrt{L_6}}exp[i(2fV_iln(cosh(\m_iz))/\m_i-m_{u,i}z)]
\label{wavesixu}
\end{equation}

\noindent
which correspond to a VEV $\phi_i(z)=V_itgh(\m_i z)$ with $\m_i = \sqrt{\l/2} V_i$.

As pointed out in Ref. \cite{PQMarcos}, $L_6$ is now a generic symbol for the length of the 
physical space, which is $L_6/2$ for the orbifold $S_1/(Z_2\times Z_2^{\prime})$. 
From Eq. (\ref{SYukeff6}) one obtains the mass matrix 

\begin{equation}
M = g_{Y,u}^{eff}\frac{v}{\sqrt{2}} \left(\begin{array}{ccc}
a_{11} & a_{12} & a_{13} \\           
a_{21} & a_{22} & a_{23} \\
a_{31} & a_{32} & a_{33} 
\end{array}\right) \, .
\label{massmat6}
\end{equation}

\noindent
Following Ref. \cite{PQMarcos}, if one now uses the linear approximation for the kink, which is 
valid for $1/\m_i \sim \cO(L_6)$, all domain wall thiknesses along $z$ are of the size of 
the compact dimension, one can obtain for the elements $a_{ij}$ the form 
$(1-\r_{ij})e^{i\theta_{ij}}$ with $\r_{ij} \ll 1$ and $\theta_{ij} \ll 1$.
In the linear approximation for the kink one obtains the following expressions for $a_{ij}$

\begin{equation}
a_{ij} = \frac{1}{L_6} \int_0^{L_6} dz \, exp [-i(\D\m_{ij}^2z^2-\D m_{ij})] \, ,
\end{equation}

\noindent
where

\begin{equation}
\D\m_{ij}^2 \equiv \frac{1}{2} (2f V_i\m_i -2f V_j \m_j) \, , 
\end{equation}
\begin{equation}
\D m_{ij} \equiv m_{q,i} -m_{u,j} \, .
\end{equation}

\noindent
As shown in Ref.  \cite{PQMarcos}, one can explicitly carry out the integration and obtain

\begin{equation}
a_{ij} = \frac{\sqrt{\pi}}{2}
\frac{ erf\left(\frac{i(2\D\m_{ij}^2L_6-\D m_{ij})}{2\sqrt{i\D\m_{ij}^2}}\right) + 
erf\left(\frac{i\D m_{ij}}{2\sqrt{i\D\m_{ij}^2}}\right)}{\sqrt{i\D\m_{ij}^2}L_6} 
exp\left(i\frac{(\D m_{ij})^2}{4\D\m_{ij}^2}\right) \, .
\label{aij}
\end{equation}

\noindent
Now if $\sqrt{\D\m_{ij}^2}L_6 \equiv x_ij \ll 1$ and $\D m_{ij} \equiv \ll 1$ one can expand 
Eq. (\ref{aij}) giving 

\begin{equation}
a_{ij} = \left(1-\frac{2}{45}x_{ij}^4-\frac{1}{24}y_{ij}^2 +\frac{1}{12} x_{ij}^2y_{ij}\right)
exp\left(i\left(\frac{y_{ij}}{2}-\frac{x_{ij}^2}{3}\right)\right) + \cO(x_{ij}^8,y_{ij}^4)\, ,
\label{almostaij}
\end{equation}

\noindent
which has the desired almost-PPMM form. 
It has to be stressed here that the expression for the mass matrix which has been used to make 
our numerical simulation is the one from Eq. (\ref{aijused}). This implies that when we 
looked for a solution in the parameter space, we did not have to restrict ourselves to the 
particular range of values for the parameters where both the linear approximation for the kink 
and Eq. (\ref{almostaij}) were valid.   

As pointed out in Ref. \cite{PQMarcos}, by looking at Eqs. (\ref{massmat6}) and (\ref{aijused}) 
one can make the important following consideration. 
If one introduces the same ``mass'' term for  $left$ and $right$ components for each family $i$, 
which means that $m_{q,i} = m_{u,i}$ (and similarly 
$m_{q,i} = m_{d,i}$ for the Down sector), 
then the mass matrix of Eq. (\ref{massmat6}) is {\em hermitian}, i.e.
$a_{ij} = a^{*}_{ji}$.
In this particular case the mass matrices for the Up and Down sectors differ only by the 
Yukawa couplings and one will not be able to reproduce a realistic mass spectrum. Moreover 
the diagonalization matrices are the same, i.e. $V_U=V_D$, and $V_{CKM}=V_U^{\dag}V_D$  
becomes just the unit matrix.    
So in order to obtain a realistic mass spectrum and CKM matrix one needs to 
introduce different ``mass'' terms for $left$ and $right$ components at least for one sector. 
What will be shown in the following is that we will be able to reproduce the right mass 
spectrum and the right CKM matrix for the case in which both up and down mass matrices are not 
hermitian. 
While hermitian mass matrices do not give a realistic 
scenario, they have the important property of having a real determinant. 
This implies that the argument of their determinant is zero. 
As pointed out in \cite{PQMarcos}, this fact could form the seed for a solution to
the strong CP problem.

\section{Description of the Parameter Space}

In this section we are going to describe the parameter space for the model of 
Ref. \cite{PQMarcos}. The particular case we consider has 10 
parameters. What has to be said  here is that  
we started our numerical simulation considering cases with a higher number of parameters and 
only after examining the results obtained, we were able to reduce the parameter space to 10.
This, by the way, is the same number of parameters found in the quark sector: Six masses
and four CKM parameters.
The 10 parameters we are considering are the following: 

\begin{itemize}
\item{$g_{Y,u}^{eff}$ and  $g_{Y,d}^{eff}$ defined 
respectively by Eq. (\ref{gYu}) and (\ref{gYd})}
\item{$\mu_1$, $\mu_2$ and  $\mu_3$ whose inverses  
give the thickness of the domain walls $\phi_1(z)$, $\phi_2(z)$ and $\phi_3(z)$ of Eq. (\ref{thickz})}
\item{$\e_{q,1}=m_{q,1}\sqrt{\l/2}/2f$ and $\e_{q,2}= m_{q,2}\sqrt{\l/2}/2f$ with $m_{q,1}$ 
the ``mass'' term for the $1^{st}$ family left component and $m_{q,2}$ for the $2^{nd}$ and 
$3^{rd}$ family left components}
\item{$\e_{u,2}=m_{u,2}\sqrt{\l/2}/2f$ and 
$\e_{u,3}=m_{u,3}\sqrt{\l/2}/2f$ the ``mass'' term respectively for the $2^{nd}$ and 
$3^{rd}$ family right components, $\e_{u,1}=0$}
\item{$\D\e$ being the common split in 
$m\sqrt{\l/2}/2f$ of $\e_{d,i}=\e_{u,i}-\D\e$ with respect to $\e_{u,i}$}
\end{itemize} 

As can be seen from the particular choice of parameters the background scalar potential 
does not break the left-right symmetry  $ \phi^q_i = \phi^u_i = \phi^d_i$ 
but on the other hand breaks the family symmetry $\phi_i \neq \phi_j$. 
The left-right symmetry is broken by choosing $\e_{q,i} \neq \e_{u,d,i}$ and by choosing 
different $\e_i$ for different indices $i$ one breaks additionally the family symmetry.   
The choice $\e_{q,3}=\e_{q,2}$ comes from the analysis of the same model in the case of 
11 parameters, where all the  $\e_{q,i}$ were different, and which gave as a result 
$\e_{q,3} \simeq \e_{q,2}$.

\section{Results for Mass Matrices from Six Dimensions}

In this section we will present the results obtained for the 
parameter space and for the quantities of Table 1.

We should point out that each solution that we found corresponds to a point in
the parameter space with all the fitted quantities of Table 1 being in the
experimental range \cite{Buras,PDG}.
In Figs.  1,2 and 3 we give the masses of the 6 quarks in $GeV$ evaluated at the $M_Z$ scale for 
three different cases corresponding to three different ranges of 
$arg(det M) = arg(det M_u) + arg(det M_d)$. 
It will be clear in the next section why it is interesting to look at the quantity $arg(det M)$
when we present a scenario for a possible solution to the Strong CP problem. 
We choose to evaluate the running masses $m_q(\mu)$ at the scale $\mu=M_Z$, because the CKM matrix 
parameters $V^{CKM}_{ij}$ are given at $\mu=M_Z$. This is a common approach for quark mass matrix 
phenomenology. See Ref. \cite{Fusaoka} for a review of the running masses and the renormalization 
group equation that describes the evolution of the running quark masses $m_q(\mu)$ with the scale 
$\mu$. The edges of each box in Figs. 1-3 give the uncertanties for the masses, which 
depend not only on the 
errors of the input parameters for the renormalization group (RG) equation, but also on 
the error of the parameter which governs 
the flow itself, i.e. the strong coupling $\a_s(M_Z)$.    

In the following we present two numerical examples corresponding respectively to $arg(det M) \sim O(1)$ and 
$arg(det M) < 2\times 10^{-10}$.
For each numerical examples we will also give the corresponding parameter space. 
Below we rewrite the expression for the mass matrix to make clear the role of each parameter.

\begin{eqnarray}
M_{ij} & = & g_{Yeff}\frac{v}{\sqrt{2}} \int dz \xi_{6,q}^{i*}(z) \xi_{6,u^c}^j(z) \nonumber \\
& = & g_{Yeff}\frac{v}{\sqrt{2}} \,\frac{1}{L_6} \,exp\left(-i\,2f\sqrt{\frac{\l}{2}}\right)\int_0^{L_6} dz \,
exp[-i(ln(cosh(\m_iz)) - ln(cosh(\m_jz)) - (\e_{q,i} - \e_{u,j})z)] \, ,\nonumber \\
\label{aijused2}
\end{eqnarray}

\noindent 
where $\e_{q,u,i} = m_{q,u,i} \sqrt{\l/2}/2f$ with  $\m_i = \sqrt{\l/2} V_i$. 
In our numerical simulation we will ignore the phase factor $exp(-i\,2f\sqrt{\frac{\l}{2}})$ 
in Eq. \ref{aijused2}, which is independent of the indices 
$i,j$.  
All results correspond to $L_6=1$. 

One word of caution is in order here. In comparing our results with the phenomenological
extractions of the CKM matrix elements, we take into account the following points.
1) The magnitudes of $V_{ub}$ and $V_{cb}$ are obtained from {\em tree-level} decays
and are to a very good approximation independent of contributions from
{\em new physics}. 2) If one were to use the Unitarity Triangle parameters
$\bar{\rho}$ and $\bar{\eta}$ for comparison, one has to assume that possible
new physics contributions (through loop effects for example) conspire to bring
the apex $(\bar{\rho},\bar{\eta})$ of the Unitarity Triangle to within the
allowed band of the ``unitarity clock''. In the following we will use both
points 1 and 2 to make our comparisons with experiments.

First, we give a numerical example corresponding to the case
$arg(detM) \sim O(1)$.


\noindent
In Eq.s (\ref{parmu})-(\ref{pargY}) we give the parameters space for the first case. 

\begin{equation}
\m_1 = 7.378    \,,\,\,\,\, \m_2=8.460    \,,\,\,\,\, \m_3=8.531     \, ,
\label{parmu}
\end{equation}
\begin{equation}
\e_{q1} =-8.262     \,,\,\,\,\, \e_{q2}=5.090    \,,\,\,\,\, \e_{q3}=5.090      \, ,
\label{parepsq}
\end{equation}
\begin{equation}
\e_{u1} = 0.000\,,\,\,\,\, \e_{u2}=1.120    \,,\,\,\,\, \e_{u3}= 1.198     \, ,
\label{parepsu}
\end{equation}
\begin{equation}
\e_{d1} = -2.044    \,,\,\,\,\, \e_{d2}=-0.924   \,,\,\,\,\, \e_{d3}= -0.846   \, ,
\label{parepsd}
\end{equation}
\begin{equation}
g_{Yu}v/\sqrt{2} =152.31 \,,\,\,\,\, g_{Yd}v/\sqrt{2}=24.48 \, .
\label{pargY}
\end{equation}

\noindent
We have decided to present the parameter space in a more readable way but it is important to remember that 
the number of independent parameters is 10.
In Eqs. (\ref{matrixup}) and (\ref{matrixdown}) we give the numerical expressions for the up and down quark 
mass matrix for the first case and in Eq.'s (\ref{eigenup}) and (\ref{eigendown}) 
the corresponding mass eigenvalues. 
As one can see, the elements of the mass matrices, although not equal to
each other in magnitudes, are of the same order, revealing their democratic origins.

\begin{equation}
M_u=152.31\,GeV\,\left(\begin{array}{ccc}
0.1111 + 0.1690i &-0.1937 - 0.4167i  &-0.1852 - 0.4377i  \\ 
0.1039 + 0.1750i &-0.1857 - 0.4221i  &-0.1765 - 0.4430i  \\
0.1031 + 0.1758i & -0.1846 - 0.4238i & -0.1753 - 0.4446i
\end{array}\right) \, ,
\label{matrixup}
\end{equation}  
\begin{equation}
m_u = 0.0023 \,GeV\,,\,\,\,\, m_c=0.624 \,GeV\,,\,\,\,\, m_t=183.0 \,GeV\, ,
\label{eigenup}
\end{equation}
\begin{equation}
M_d= 24.48\, GeV \,\left(\begin{array}{ccc}
-0.0105 + 0.0003i &-0.0424 - 0.0086i  & -0.0544 - 0.0121i \\ 
-0.0090 + 0.0041i & -0.0442 - 0.0060i & -0.0563 - 0.0099i \\
-0.0081 + 0.0043i &-0.0451 - 0.0060i  & -0.0572 - 0.0100i
\end{array}\right) \, ,
\label{matrixdown}
\end{equation}  
\begin{equation}
m_d = 0.0048 \,GeV\,,\,\,\,\, m_s=0.081 \,GeV\,,\,\,\,\, m_b=3.09 \,GeV\, .
\label{eigendown}
\end{equation}

In Eqs. (\ref{absmatup}) and (\ref{absmatdown}) we also give the absolute values of the mass matrices 
of Eqs. (\ref{matrixup}) and (\ref{matrixdown}) which show that {\bf breaking the family symmetry} 
does not destroy completely the democratic structure of the mass matrices.

\begin{equation}
M_u=152.31\,GeV\,\left(\begin{array}{ccc}
0.2023  & 0.4595 &0.4753 \\           
0.2035  &0.4611 & 0.4768 \\
0.2038  &0.4622  & 0.4780
\end{array}\right) \, ,
\label{absmatup}
\end{equation}  
  
\begin{equation}
M_d=24.48\,GeV\,\left(\begin{array}{ccc}
0.0105 &0.0432  & 0.0558 \\           
0.0099 &0.0446  &0.0571 \\
0.0092 & 0.0455 & 0.0581
\end{array}\right) \, .
\label{absmatdown}
\end{equation}  

In Eqs. (\ref{absmatup2}) and (\ref{absmatdown2}) we give  the absolute values of the mass matrices 
of Eqs. (\ref{matrixup}) and (\ref{matrixdown}) with rescaled values of the matrix elements. Eqs.  
(\ref{absmatup2}) and (\ref{absmatdown2}) show in a more explicit way that the deviations from a 
democratic mass matrix are of $O(1)$.

\begin{equation}
M_u=72.80\,GeV\,\left(\begin{array}{ccc}
0.4232 &0.9614  &0.9943  \\           
0.4257 & 0.9647 &0.9976 \\
0.4264 &0.9671  &  1.0000 
\end{array}\right) \, ,
\label{absmatup2}
\end{equation}  

\begin{equation}
M_d=1.42\,GeV\,\left(\begin{array}{ccc}
0.1807 & 0.7446 &0.9601  \\           
0.1702 &0.7675  &0.9836 \\
0.1584 &0.7836  &  1.0000
\end{array}\right) \, .
\label{absmatdown2}
\end{equation}  

\noindent
In Eq. (\ref{CKMmat}) we give the CKM matrix corresponding to the mass matrices of Eqs. 
\ref{matrixup} and (\ref{matrixdown}), in Eq. (\ref{absCKMmat}) its absolute value, and 
in Eqs. (\ref{rhoetaval}) and (\ref{betagamval}) the values for the parameters $\bar{\rho}$ 
and $\bar{\eta}$, and for $\sin{2\b}$ and $\g$.

\begin{equation}
V_{CKM}=\left(\begin{array}{ccc}
0.9711 - 0.0884i &0.0930 - 0.2014i   &-0.0036 + 0.0010i  \\           
-0.1185 - 0.1872i &0.9734 - 0.0392i  &-0.0144 + 0.0398i  \\
0.0095 - 0.0064i & 0.0150 + 0.0380i &  0.9986 + 0.0302i
\end{array}\right) \, ,
\label{CKMmat}
\end{equation}  
   
\begin{equation}
V_{CKM}=\left(\begin{array}{ccc}
0.9751  &0.2218  &0.0038  \\           
0.2215 & 0.9742  & 0.0423 \\
0.0423 &0.0409   & 0.9991
\end{array}\right) \, ,
\label{absCKMmat}
\end{equation}  

\begin{equation}
\bar{\rho} =0.18 \, , \,\,\,\,\, \bar{\eta} =0.35 \, .
\label{rhoetaval}
\end{equation}
\begin{equation}
\sin{2\b} =0.72 \, , \,\,\,\,\, \g =63.2^0 \, .
\label{betagamval}
\end{equation}

\noindent
with $\bar{\rho}$ and $\bar{\eta}$ being defined as

\begin{equation}
\bar{\rho} = Re(V_{ud}V_{ub}^*V_{cd}^*V_{cb})/|V_{cd}V_{cb}^*|^2 \, , 
\end{equation}
\begin{equation}
\bar{\eta} = Im(V_{ud}V_{ub}^*V_{cd}^*V_{cb})/|V_{cd}V_{cb}^*|^2 \, .
\end{equation}

\noindent
and $\sin{2\b}$ and $\g$ as

\begin{equation}
\sin{2\b} = \frac{2\bar{\eta}(1-\bar{\rho})}{(1-\bar{\rho})^2+\bar{\eta}^2}
\end{equation}
\begin{equation}
\g=90^0-\frac{90^0}{\pi}sin^{-1}\left(\frac{2\bar{\rho}\bar{\eta}}{\bar{\rho}^2+\bar{\eta}^2}\right) 
\end{equation}

In Eq. (\ref{argdetval}) we give the values for the $arg(det M_u)$, $arg(det M_d)$ and for their sum.

\begin{equation}
arg(det M_u) = -1.5692 \, , \,\, arg(det M_d) =1.8643  \, , \,\,arg(det M)= 0.2951\,  .
\label{argdetval}
\end{equation}


The following numerical results corresponding  to a value of  $arg(det M) < 10^{-9}$ are presented 
in the same way as the above example.

\begin{equation}
\m_1 = 7.365\,,\,\,\,\, \m_2=8.456\,,\,\,\,\, \m_3=8.532 \, ,
\label{parmu_2}
\end{equation}
\begin{equation}
\e_{q1} = -8.189\,,\,\,\,\, \e_{q2}=5.026\,,\,\,\,\, \e_{q3}=5.026 \, ,
\label{parepsq_2}
\end{equation}
\begin{equation}
\e_{u1} =0.000 \,,\,\,\,\, \e_{u2}=1.105\,,\,\,\,\, \e_{u3}=1.188\, ,
\label{parepsu_2}
\end{equation}
\begin{equation}
\e_{d1} =-2.274 \,,\,\,\,\, \e_{d2}=-1.168\,,\,\,\,\, \e_{d3}= -1.085\, ,
\label{parepsd_2}
\end{equation}
\begin{equation}
g_{Yu}v/\sqrt{2} =144.63 \,,\,\,\,\, g_{Yd}v/\sqrt{2}=23.60\, .
\label{pargY_2}
\end{equation}

\begin{equation}
M_u=144.63\,GeV\,\left(\begin{array}{ccc}
0.1154 + 0.1622i & -0.1846 - 0.4383i & -0.1740 - 0.4608i \\    
0.1096 + 0.1662i & -0.179 - 0.4367i & -0.1683 - 0.4591i \\
0.1090 + 0.1670i & -0.1781 - 0.4382i &  -0.1671 - 0.4606i
\end{array}\right) \, ,
\label{matrixup_2}
\end{equation}  
\begin{equation}
m_u = 0.0028\,GeV\,,\,\,\,\, m_c=0.621 \,GeV\,,\,\,\,\, m_t=177.9 \,GeV\, ,
\label{eigenup_2}
\end{equation}

\begin{equation}
M_d= 23.60\, GeV \,\left(\begin{array}{ccc}
-0.0609 + 0.0113i & -0.0167 - 0.0046i & -0.0294 - 0.0063i \\         
-0.0630 + 0.01480i &-0.0144 - 0.0006i  & -0.0272 - 0.0025i \\
-0.0622 + 0.0147i & -0.0152 - 0.0005i &  -0.0280 - 0.0024i
\end{array}\right) \, ,
\label{matrixdown_2}
\end{equation}  
\begin{equation}
m_d = 0.0047 \,GeV\,,\,\,\,\, m_s= 0.105\,GeV\,,\,\,\,\, m_b=2.9 \,GeV\, .
\label{eigendown_2}
\end{equation}

\begin{equation}
M_u=144.63\,GeV\,\left(\begin{array}{ccc}
0.1990 & 0.4756 & 0.4925 \\           
0.1991 &0.4720  &0.4889  \\
0.1994 & 0.4730 &0.4899
\end{array}\right) \, ,
\label{absmatup_2}
\end{equation}  

\begin{equation}
M_d=23.60\,GeV\,\left(\begin{array}{ccc}
0.0619 &0.0173 &0.0301  \\           
0.0647 &0.0144  & 0.0273 \\
0.0639 &0.0152  & 0.0281 
\end{array}\right) \, .
\label{absmatdown_2}
\end{equation}

\begin{equation}
M_u=71.23\,GeV\,\left(\begin{array}{ccc}
0.4042 &0.9656  & 1.000 \\           
0.4043 & 0.9583 & 0.9927 \\
0.4049 &0.9604  & 0.9948 
\end{array}\right) \, ,
\label{absmatup2_2}
\end{equation}  

\begin{equation}
M_d=1.53\,GeV\,\left(\begin{array}{ccc}
0.9568 &0.2676 &0.4655  \\           
1.0000 & 0.2229 & 0.4223 \\
0.9873 &0.2350  & 0.4346 
\end{array}\right) \, .
\label{absmatdown2_2}
\end{equation}

\begin{equation}
V_{CKM}=\left(\begin{array}{ccc}
-0.9681 + 0.1152i &-0.1707 + 0.1425i & 0.0038 + 0.0011i \\   
-0.1879 - 0.1183 &0.9741 - 0.01456i  & -0.0078 + 0.0384i \\
0.0068 - 0.0096i & 0.0082 + 0.0367i & 0.9991 + 0.0126i
\end{array}\right) \, ,
\label{CKMmat_2}
\end{equation}  
   
\begin{equation}
V_{CKM}=\left(\begin{array}{ccc}
0.9750 & 0.2223 &0.0040  \\           
0.2220 & 0.9743 & 0.0392 \\
0.0118 & 0.0376 & 0.9992 
\end{array}\right) \, ,
\label{absCKMmat_2}
\end{equation}  

\begin{equation}
\bar{\rho} = 0.31 \, , \,\,\,\,\, \bar{\eta} = 0.32\, ,
\label{rhoetaval_2}
\end{equation}
\begin{equation}
\sin{2\b} =0.77 \, , \,\,\,\,\, \g = 46.3^0\, ,
\label{betagamval_2}
\end{equation}

\begin{equation}
arg(det M_u) = -1.55845421528\, , \,\, arg(det M_d) =1.55845421485  \, , \,\,arg(det M)=4.3\times 10^{-10} \,  .
\label{argdetval_2}
\end{equation}

\noindent
In Figs. 4, 5 and 6 we give the absolute values of the CKM matrix elements for the three 
different cases corresponding 
to three different ranges of $arg(det M)$. 
The uncertainties for each element are given by the edges of the corresponding window.   
Fig. 8 shows the solutions for $\bar{\rho}$ and $\bar{\eta}$. The sharp edges for the 
solution patches are due 
to the constraints imposed on $\bar{\rho}$ and $\bar{\eta}$.  
Fig. 9 shows instead the solutions for $\sin{2\b}$ and $\g$. 
It  has to be said here that the solutions appear in patches because of the way the minimization 
process works. 
One obviously can not exclude other solution patches. For example, could we try to make the 
minimization process follow 
different paths by toying with the input parameters, temperature and number of iterations, 
(see appendix A), and by changing the initial conditions.

\begin{table}
\begin{center}
\begin{tabular}{ccc}
$x_i$ & $<x_i>$ & $|x^{max}_i-x^{min}_i|/2$ \\ \hline \\
$m_u$ & $2.33 \times 10^{-3}$ & $0.45 \times 10^{-3}$ \\ 
$m_c$ & $0.685 $ & $0.061$ \\ 
$m_t$ & $181$ & $13$ \\ 
$m_d$ & $4.69 \times 10^{-3}$ & $0.66 \times 10^{-3}$ \\ 
$m_s$ & $0.0934 $ & $0.0130$ \\ 
$m_b$ & $3.00$ & $0.11$ \\ 
$m_u/m_d$ & $0.497$ & $0.119$ \\ 
$m_s/m_d$ & $19.9$ & $3.9$ \\ 
$|V_{ud}|$ & $0.97485$  & $0.00075$ \\ 
$|V_{us}|$ & $0.2225$  & $0.0035$ \\ 
$|V_{ub}|$ & $0.00365$  & $0.00115$ \\ 
$|V_{cd}|$ & $0.2225$  & $0.0035$ \\ 
$|V_{cs}|$ & $0.9740$  & $0.0008$ \\ 
$|V_{cb}|$ & $0.041$  & $0.003$ \\ 
$|V_{td}|$ & $0.009$  & $0.005$ \\ 
$|V_{ts}|$ & $0.0405$  & $0.0035$ \\ 
$|V_{tb}|$ & $0.99915$  & $0.00015$ \\ 
$\bar{\rho}$ & $0.22$ & $0.10$ \\ 
$\bar{\eta}$ & $0.35$ & $0.05$  \\ 
$\sin{2\b}$ & $0.78$ & $0.08$ \\
$\g$ & $59^0$ & $13^0$ \\ 
\end{tabular}
\caption{Central values and uncertainties for the masses of the 6 quarks 
evaluated at $M_Z$, for 
the two ratios $m_u/m_d$ and $m_s/m_d$, for the absolute values of the CKM matrix elements, 
the CP parameteres $\bar{\rho}, \bar{\eta}$, $\sin{2\b}$ and $\g$.  }
\end{center}
\end{table}

The parameter space with the 10 parameters described in the previous section and  which corresponds to 
the three different cases for the the three different ranges of $ arg(det M)$, are given in 
Figs. 10, 11 and 12. 
It has to be said here that the solutions presented correspond to a two step procedure. 
First we find solutions by requiring $\m_1$, $\m_2$ and $\m_3$ to be larger than unity, because 
one can use in the physical space $[0,L_6]$ the kink solution for $\phi_i(z)$ instead of the 
kink-antikink approximate solution $Vtgh(\mu z)tgh(\mu(L_6-z))$ \cite{Georgi} (the kink-antikink solution 
has the property of vanishing at both orbifold fixed point $z=0$ and $z=L_6$ as required by the 
imposed boundary conditions to compactify on an $S_1/Z_2$ orbifold).
Second, using initial conditions from the parameter space already found, we looked for solutions
whic correspond to very small ranges of $\m_1$, $\m_2$ and $\m_3$. The reason we did this 
is because we noticed 
that the parameters which were more relevant to fit the quantities of Table 1  were the 
``mass'' terms of left and right components, i.e $\e_{q,i}$, $\e_{u,i}$ and $\e_{d,i}$ respectively.
As a result, in order to understand the dependence of the found solutions on the ``mass'' terms, 
we decided to restrict the range of $\m_1$, $\m_2$ and $\m_3$. 
The way one controls a range for a parameter consists 
simply in adding that parameter to the quantities one wants to fit, modifying the function $f$ of 
Eq. (\ref{ff}) given in appendix A.
For our numerical simulation we set $L_6=1$. This implies that the introduced parameters 
$\mu$ and $\e$ have to be multipied by $L_6$ for a general case.

We have presented in this section an analysis of the quark mass matrices obtained in
Ref. \cite{PQMarcos}. In this analysis, we have used 10 parameters and were successful
in fitting all 6 quark masses and all the parameters of the CKM matrix. This is shown in
Figs. (1-10) for three separate ranges of the quantity $|arg(detM)|$ which was defined
above. In these figures, each dot in the scatter plot represents a solution which fits
the masses and the CKM matrix. As one can see, the solutions which correspond to
$|arg(detM)| \sim 10^{-10}$ and which will have a significance to the Strong CP
problem, are within $1\, \sigma$ and $1.5 \, \sigma$ of the so-called R-fit and
Bayesian fit respectively, as presented in Ref. \cite{Buras}.

   

\section{Possible connection between Strong and Weak CP}

In this section, we will discuss a possible connection between the region of parameter space
where $ arg(det M) < 10^{-10}$ and the Strong CP problem.

It is well-known that the non-trivial vacuum of QCD generates a P and CP 
violating term in the Lagrangian of the form

\begin{equation}
{\cal L}_{\bar{\theta}} = \frac{\bar{\theta}}{32 \pi^2}\tilde{G}_{\mu\nu}G^{\mu\nu}
\label{Ltheta}
\end{equation}
where
\begin{equation}
\label{thetabar}
\bar{\theta}=\theta_{QCD} + arg(det M)\,.
\end{equation}
This term (\ref{Ltheta}) gives a contribution to the electric dipole moment of the neutron
\cite{Baluni}, \cite{Crewther}, with
the current experimental limit \cite{Barnett} being
$\bar{\theta}<2\times10^{-10}$. 
The mystery of why $\bar{\theta}$ should be so small constitutes the so-called Strong CP problem. 

The most famous and elegant solution to the Strong CP problem is the Peccei-Quinn mechanism
\cite{PecceiQuinn} where $\bar{\theta}$ becomes a dynamical field and
relaxes to zero at the minimum of its potential. This dynamical field manifests itself
as a pseudoscalar particle- the so-called Axion- whose decay constant is now severely
constrained by astrophysical and cosmological arguments \cite{axion}. Another
solution involving a massless up quark \cite{masslessup} is largely disfavoured
by studies of chiral perturbation theory. A third type of solution to the 
Strong CP problem which has no axion, is the Nelson-Barr type of mechanism
\cite{nelsonbarr} which
assumes exact CP at tree level and whose mass matrices have
real determinants, and, as a result, the strong CP problem does not exist. It can arise
at loop levels and can be ``under control''. However a realistic model of this type
is yet to be constructed.



We have already mentioned that when we introduce the same ``mass'' term for left and right components 
(Eqs. (\ref{massmat6}) and (\ref{aijused})), i.e. 
$m_{q,i}=m_{u,i}=m_{d,i}$, the up and down quark mass matrices are 
hermitian and consequently their determinant is real \cite{PQMarcos}.  $arg(detM_u)$ and $arg(detM_d)$ are 
separately zero. This situation suggests that the symmetry of the Lagrangian that makes 
$arg(detM)=0$ at tree level is the ``left-right'' symmetry of the components along 
the sixth dimension. If we also assume CP conservation at the Lagrangian level, this scenario
would provide a solution to the strong CP problem.
However this symmetry has to be broken because, as we have already mentioned, the case 
in which the quark mass matrices are hermitian does not reproduce the right mass spectrum and 
the right CKM matrix. One could imagine a scenario where the ``left-right'' symmetry is spontaneously 
broken. This will induce some loop corrections that will make the mass matrices deviate from 
hermiticity, reproducing the right mass spectrum and right CKM matrix. At the same time the 
breaking of this ``left-right'' symmetry could induce at loop levels a nonvanishing $\bar{\theta}$. 
But one can envision a scenario were the deviation from hermiticity happens in such a way that 
$arg(detM_u)$ and $arg(detM_d)$, each being now of $O(1)$, can cancel each other so as to keep  
$\bar{\theta}<2\times10^{-10}$. 

As we have mentioned earlier, the solutions presented in this paper correspond to 
very small windows for the domain wall parameters $\m_i$'s, because we wanted 
to put in evidence the effect of the ``left-right'' 
symmetry breaking, especially in the $\bar{\theta}$ parameter.
In searching for a quantity which could ``retain the memory'' of this 
``left-right'' symmetry, we decided 
to plot, see Fig. 13, the sum of the arguments $|arg(det(M))|$ versus the ``CoM'',
the weighted average of the ``mass'' terms along the sixth dimension, which is defined as

\begin{equation}
CoM \equiv \frac{2\sum_i\e_{q,i}+\sum_i\e_{u,i}+\sum_i\e_{d,i}}{12} \, .
\label{CoM}
\end{equation}

\noindent
As one can notice from Fig. 13, the sum of the arguments tends to go to zero for 
a particular $CoM \simeq 0.125$, and 
this behavior is confirmed in  Fig. 14. Notice that, for practical reasons, we have
reduced the number of points (i.e. the number of solutions) in Figs. 13 and 14 in
order to reduce the sizes of the files containing these two figures. The actual
number of solutions is much larger than what is shown in these figures.

Now one can think that the value of $0.125$ for the ``CoM'' when the ``left-right'' 
symmetry is broken, 
and which corresponds to $\bar{\theta}<2\times10^{-10}$, was also the value for the ``CoM'' 
before the breaking, when $\bar{\theta}$ was equal to zero. 
In other words one can imagine a scenario where the ``mass'' terms of left and right components 
are split such a way as to retain the same value of the ``CoM''. 
To invent a mechanism which could break the ``left-right'' symmetry and 
which could reproduce the scenario mentioned 
above is beyond the scope of this paper, but it is one of the main topics we would like 
to investigate in the future.    

If one now looks at Fig. 8  which show the solutions in 
the $\bar{\rho}, \bar{\eta}$ plane for $|arg(det M)|$ in the three different ranges, one can see 
that the solutions corresponding to $10^{-12}<|arg(det M)|<2 \times 10^{-10}$ 
tend to favor a particular region 
of the $\bar{\rho}, \bar{\eta}$ plane. 
In particular, the region of allowed solutions shrinks down in the $\bar{\rho}$ 
direction constraining the parameter 
$\bar{\rho}$  in the small window $\sim [0.3, 0.32]$. 
These solutions are within $1\, \sigma$ and $1.5 \, \sigma$ of the so-called R-fit and
Bayesian fit respectively as can be seen from Fig. 8.    
This could be an artifact of the way the minimization process works, but it could also be an 
indication that there is a deep connection between Weak CP violation and Strong CP 
violation in our model. 

As we have already mentioned earlier, one cannot exclude completely the existence 
of other solutions, because of 
the way the minimization process works, but we believe that the results obtained here 
can give some very good indications of how the allowed parameter space might look like.

\section{Epilogue}

We have presented in this paper a complete phenomenological analysis of the model of
quark mass matrices as derived from six dimensions by \cite{PQMarcos}. With just 10
parameters, we have found a large number of solutions which can fit the 6 quark
masses as well as the CKM matrix as can be seen in
Figs. (1-9). Furthermore, we subdivide these
solutions into three sets: 1) Those that have $10^{-1} < |arg(det M)| < 5 \times 10^{-1}$,
2) those that have $10^{-3} < |arg(det M)| < 10^{-1}$, and finally 3) those that have
$10^{-12} < |arg(det M)| < 2 \times 10^{-10}$. The first two sets are given for the sole
purpose of comparison with the last set which is most relevant to the Strong CP problem
as discussed in the last section. As one can observe from Figs. (8) and (9), there is
a deep connection between the weak CP parameters and the Strong CP phase. In order to
satisfy the constraint on the $\bar{\theta}$ parameter, i.e. $|arg(det M)| < 2 \times 10^{-10}$
in our framework, the solutions obtained for $\bar{\rho}$ and $\bar{\eta}$ are found
to be within $1\, \sigma$ and $1.5 \, \sigma$ of the so-called R-fit and
Bayesian fit respectively. As measurements of weak CP parameters become more
accurate in the future, they will either rule out or confirm our predictions.

\acknowledgments
We would like to thank Andrzej Buras and Gino Isidori for valuable comments.
One of us (PQH) would like to thank the theory group of Laboratorio Nazionale di
Frascati for hospitality during the course of this work. (AS) would like to thank Mike Timmins 
for introducing the simulated annealing method and Ngoc-Khanh Tran for usefull discussions.  
Our work is supported in parts by the US Department of Energy under grant No. DE-A505-89ER40518.

\section{Appendix A} 
 
As mentioned above the model we considered has $10$ free parameters. Relying on this freedom 
we were able to fit the 6 quark masses evaluated at an energy scale equal 
to the mass of Z gauge boson $M_Z$, with constraints for the two ratios $m_u/m_d$ and $m_s/m_d$, 
the absolute values of the CKM matrix 
elements and the CP parameters $\bar{\rho}, \bar{\eta}$, or equivalently the three angles and one phase of
 the CKM matrix 
standard parametrization, for a total number of 10 quantities. 
The approach we used to derive the parameter space consists in minimizing a 
particular function, built in such a way that its global minima correspond 
to the region defined by the experimental constraints. 
This function is defined in the following way:

\begin{eqnarray}
f & = & \sum_{i=1}^N \frac{(x^{th}_i-x^{min}_i)^2}{<x_i>^2}
\,\theta(x^{min}_i-x^{th}_i) \nonumber \\
& + & \sum_{i=1}^N \frac{(x^{th}_i-x^{max}_i)^2}{<x_i>^2}
\,\theta(x^{th}_i-x^{max}_i)
\label{ff}
\end{eqnarray} 

\noindent
where $\theta(x)$ is the step function, N is the number of quantities 
that we want to fit, $x^{th}_i$ is the 
predicted value for the $ith$ quantity, $x^{min}_i $ and $x^{max}_i$ 
fix the range for the $ith$ quantity, and $<x_i>$ is its average value.   
It is immediate to verify from Eq. (\ref{ff}) that when all the predicted 
quantities $x^{th}_i$'s are contained in the proper ranges, the function $f$ 
takes its minimum value equal to zero. 

The set of parameters which correspond to a zero value for the 
function $f$ is called a solution.
In our particular case the function we are considering is a mapping from
${\cal R}^{10}$ to ${\cal R}$. The parameter space is really big and to 
find the solutions, which correspond to the global minima of $f$, can be very 
challenging. 
The global minima of $f$ can be infact surrounded by a lot of local minima 
which most of the time can make the minimization process fail. 
Therefore, just trying different initial conditions with the hope to find 
the good one that 
will allow the minimization process to follow the 
right path towards a global minimum can result quite inefficient. 

It is a common belief that there does not exist a general recipe to follow 
for minimization problems. A minimization procedure that can be very 
efficient for a particular problem, can be very inefficient in an another 
case or even fails. 
In the particular case where the function that we want to minimize 
depends on many parameters, there is a minimization procedure, called 
simulated annealing \cite{SA1}\cite{SA2}, 
which seems to work more efficiently than others. This procedure 
is mostly used when the global minima are surrounded by a lot of local minima. 
Infact this minimization process can find a global minimum also after being 
trapped in a local minimum. 
The way instead most of the minimization processes work is to go, from the 
starting point, immediately downhill as far as they can go, but this often 
leads to a local minimum. 

At the heart of the method of simulated annealing is an analogy with 
thermodynamics, specifically with the way liquids crystallize, or metal 
cool and anneal. 
When a liquid is cooled down sufficiently slowly, the atoms are often able 
to line themeselves up and form a pure crystal, which corresponds to 
the state of minimum energy for the system. 
But if the liquid is cooled quickly, it does not reach this state but rather 
ends up in an amorphous state of higher energy.
So nature, as long as the process of cooling is sufficiently slow, is able 
to find the minimum energy state. 
The way nature works is based on the fact that for a system in thermal 
equilibrium at temperature T, the probability for the system to be in a state 
of energy $E$ is given by the Boltzmann probability distribution:

\begin{equation}
Prob(E) \sim exp\left(-\frac{E}{KT}\right).
\label{Boltz}
\end{equation}

\noindent
This implies that even at low temperature there is a chance, albeit a tiny 
one, for the system to be in a high energy state. 
Therefore, there exists the possibility  for the system to get out of a local energy 
minimum and move towards the global one. 
This principle has been incorporated in what is called a Metropolis 
algorithm \cite{Metropolis}. 
Given a simulated thermodynamic system, it is assigned a 
probability $P=exp(-(E_2-E_1)/KT)$ to the change from a configuration with 
energy $E_1$ to one with energy $E_2$. If $E_2<E_1$, $P$ is greater than 
unity and in this case to the change is assigned a probability $P=1$, which 
is equivalent to say that the system always makes such a change.
In the case $E_2>E_1$, one can compare the probability $P=exp(-(E_2-E_1)/KT)$ with 
a $random-number$ and make the change to the new configuration only if $P>random-number$.   
The system always takes a downhill step while sometimes takes an uphill step. 
The Metropolis algorithm can be used for systems other than the 
thermodynamic ones if we give:

\begin{itemize}
\item A description of possible system configurations.
\item A generator of random changes in the configuration.
\item A function $E$ (analog of energy) whose minimization is the 
goal of the procedure.
\item A control parameter $T$ (analog of temperature) and an annealing 
schedule which tells how it is lowered, e.g. after how many random 
changes in configuration and with which step. 
\end{itemize}

Going back to our particular case, the function $f$ we want to minimize is 
the analog of the energy, and each possible set of parameters correspond to a 
particular configuration of the system. 
For the algorithm to work a control parameter $T$, with an annealing 
schedule by which it is gradually reduced, has to be introduced, as well 
as a generator of random changes in the configuration, that is in the 
parameter space. 
The way these random changes are taken is the following:
a positive quantity , given by $-T \cdot ln(random-number)$, 
is added to the stored function value.  
The same quantity is subtracted from the function value corresponding to 
every new set parameters that are tried as a replacement point in the 
parameter space. (The new points are obtained using the downhill simplex 
method. The simplex is a set of $N+1$ points with N the number of 
parameters, and the changes happen through reflections, expansions 
and contractions of the simplex). 
As mentioned before, this method allowed the system to jump from a local 
minimum and to look for a global one. The algorithm that we have used has 
been taken from numerical recipes \cite{NR}.
Other than the initial set of values for the parameters, also the value for 
the temperature and the number of iterations which determine the annealing 
schedule to reduce the temperature, has to be given as input. The way these 
last two values were chosen, as well as the initials conditions is mostly 
the result of different attempts.   
The output of our code is a patch of solutions, which have been subsequently 
tested using an independent code written in Mathematica.

\begin{figure} 
\caption{Solutions for the 6 quark masses corresponding to $10^{-1} < |arg(det M)| < 5 \times 10^{-1}$ . 
The masses in $GeV$ are evaluated at the $M_Z$ scale. 
The range for each mass as given in Table 1 is defined by the edges of the corresponding window.}
\end{figure}

\begin{figure} 
\caption{Solutions for the 6 quark masses corresponding to $10^{-3} < |arg(det M)| < 10^{-1}$ . 
The masses in $GeV$ are evaluated at the $M_Z$ scale. 
The range for each mass as given in Table 1 is defined by the edges of the corresponding window.}
\end{figure}

\begin{figure} 
\caption{Solutions for the 6 quark masses corresponding to $10^{-12} < |arg(det M)| < 2 \times 10^{-10}$ . 
The masses in $GeV$ are evaluated at the $M_Z$ scale. 
The range for each mass as given in Table 1 is defined by the edges of the corresponding window.}
\end{figure}

\begin{figure} 
\caption{Solutions for the absolute values of the CKM elements corresponding 
to $10^{-1} < |arg(det M)| < 5\times 10^{-1}$. 
The range for each element as given in Table 1 is defined by the edges of the corresponding window. 
The ranges delimeted by the dashed lines correspond to the new evaluations for 
$|V_{us}|$ (region on the left of the dashed line), $ |V_{ub}|$ and $V_{cb}$ as in Ref. 4.}
\end{figure}

\begin{figure} 
\caption{Solutions for the absolute values of the CKM elements corresponding 
to $10^{-3} < |arg(det M)| < 10^{-1}$. 
The range for each element as given in Table 1 is defined by the edges of the corresponding window. 
The ranges delimeted by the dashed lines correspond to the new evaluations for 
$|V_{us}|$ (region on the left of the dashed line), $ |V_{ub}|$ and $V_{cb}$ as in Ref. 4.}
\end{figure}

\begin{figure} 
\caption{Solutions for the absolute values of the CKM elements corresponding 
to $10^{-12} < |arg(det M)| < 2 \times 10^{-10}$. 
The range for each element as given in Table 1 is defined by the edges of the corresponding window. 
The ranges delimeted by the dashed lines correspond to the new evaluations for 
$|V_{us}|$ (region on the left of the dashed line), $ |V_{ub}|$ and $V_{cb}$ as in Ref. 4.}
\end{figure}

\begin{figure}
\caption{Solutions for the ratio $|V_{ub}| / |V_{cb}|$ for the three cases corresponding to a) 
$10^{-1} < |arg(det M)| < 5 \times 10^{-1}$, b) 
$10^{-3} < |arg(det M)| < 10^{-1}$,  and 
c) $10^{-12} < |arg(det M)| < 2 \times 10^{-10}$.}
\end{figure}


\begin{figure}
\caption{Solutions for $\bar{\rho}$ and $\bar{\eta}$ for the three cases, from top to bottom, 
corresponding to a) $10^{-1} < |arg(det M)| < 5 \times 10^{-1}$, b) 
$10^{-3} < |arg(det M)| < 10^{-1}$,  and 
c) $10^{-12} < |arg(det M)| < 2 \times 10^{-10}$. 
The delimeted area is the allowed region in the 
$\bar{\rho},\bar{\eta}$ plane as obtained from Ref. 4. }
\end{figure}

\begin{figure}
\caption{Solutions for $\sin{2\b}$ and $\g$ for the three cases, from top to bottom, 
corresponding to a) $10^{-1} < |arg(det M)| < 5 \times 10^{-1}$, b) 
$10^{-3} < |arg(det M)| < 10^{-1}$,  and 
c) $10^{-12} < |arg(det M)| < 2 \times 10^{-10}$.}
\end{figure}

\begin{figure}
\caption{Summary of the 10 parameters space corresponding to 
$10^{-1} < |arg(det M)| < 5 \times 10^{-1}$.}
\end{figure}

\begin{figure}
\caption{Summary of the 10 parameters space corresponding to 
$10^{-3} < |arg(det M)| < 10^{-1}$.}
\end{figure}

\begin{figure}
\caption{Summary of the 10 parameters space corresponding to 
$10^{-12}$ $<|arg(det M)|<$ $2\times10^{-10}$.}
\end{figure}

\begin{figure}
\caption{Solutions for $|arg(det M)|$ in the region 
$[10^{-3}, 5 \times 10^{-1}]$ vs the CoM as defined by Eq. \ref{CoM}.}
\end{figure}

\begin{figure}
\caption{Solutions for $|arg(det M)|$ in the region 
$[10^{-12}, 2 \times 10^{-10}]$ vs the CoM as defined by Eq. \ref{CoM}.}
\end{figure}

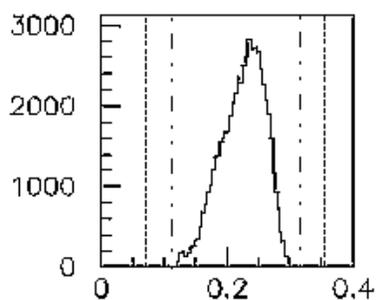
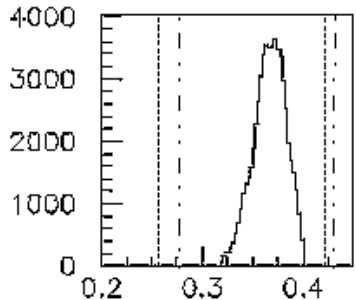
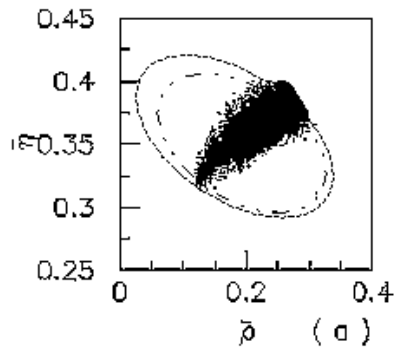
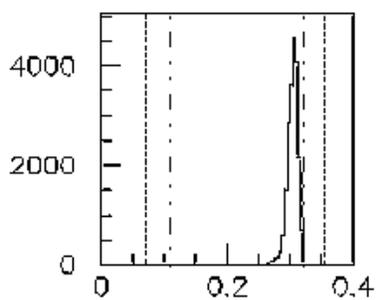
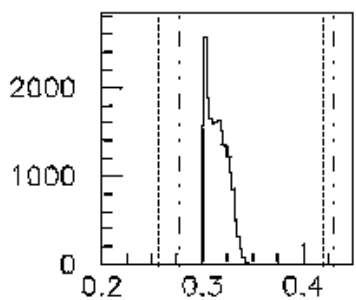
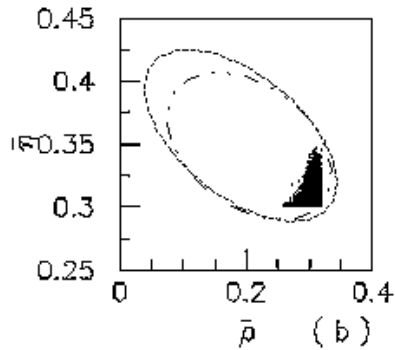
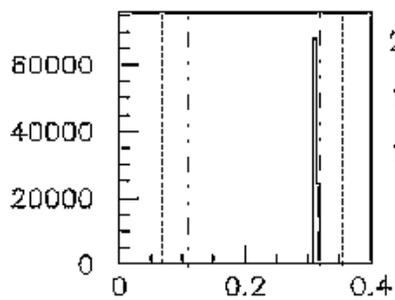
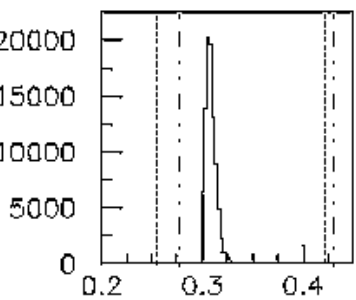
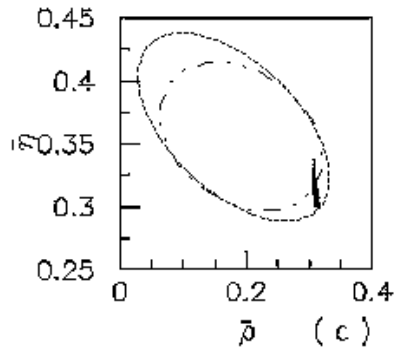
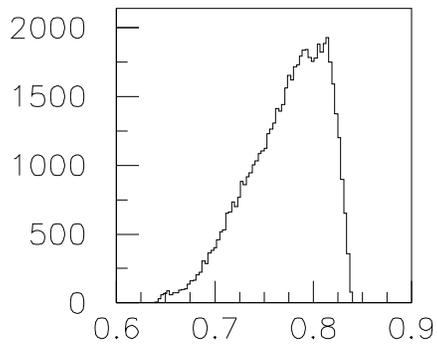
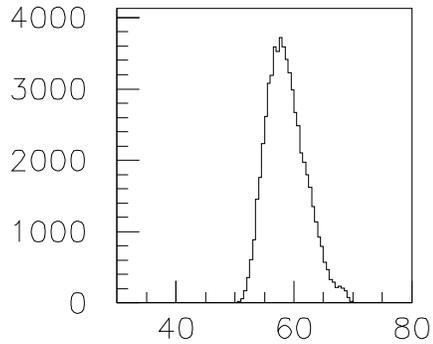
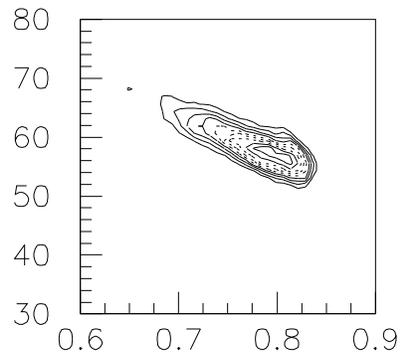
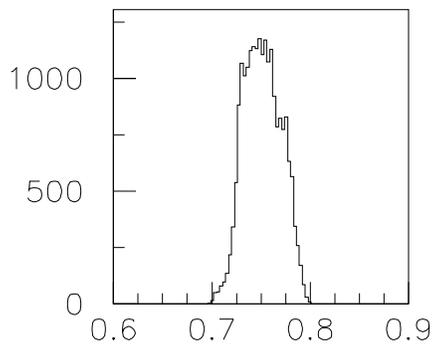
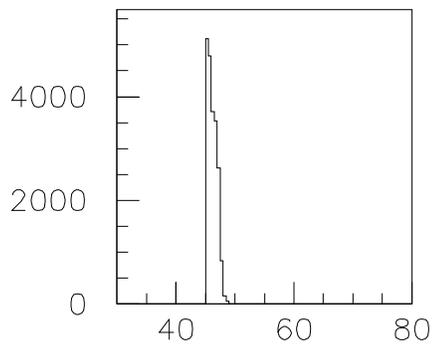
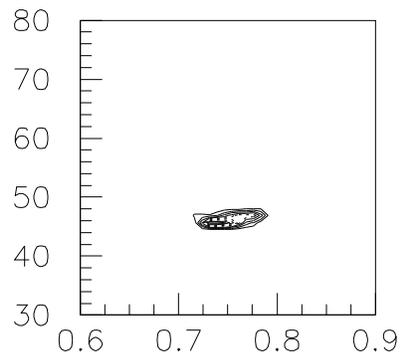
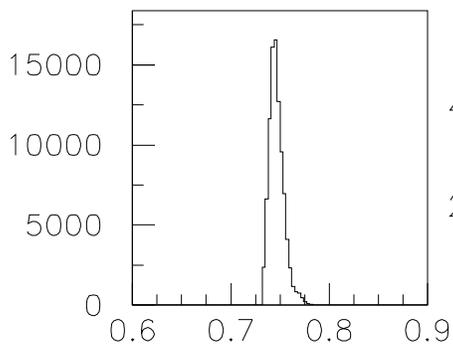
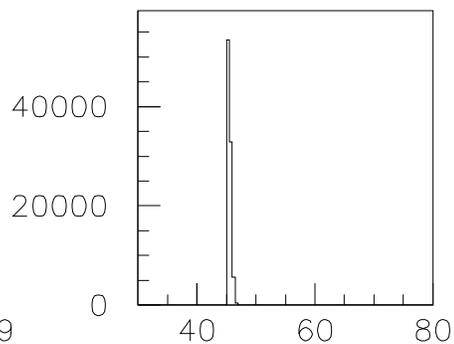
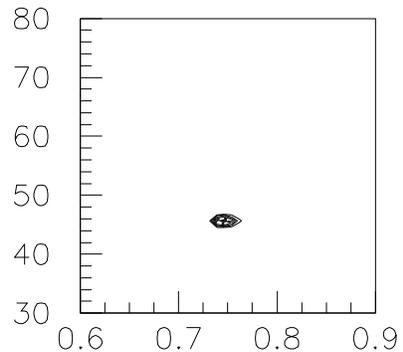
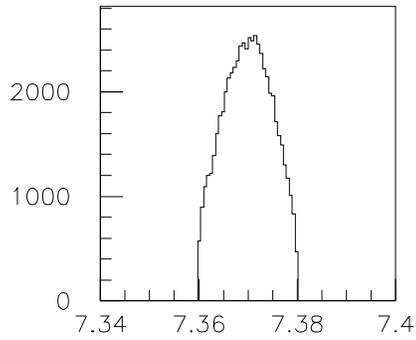
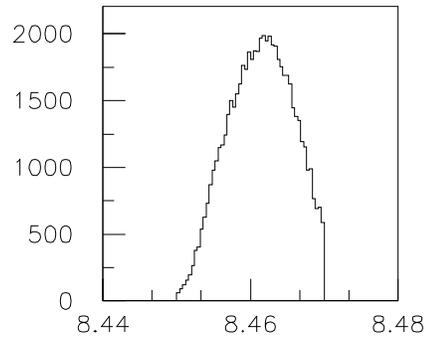
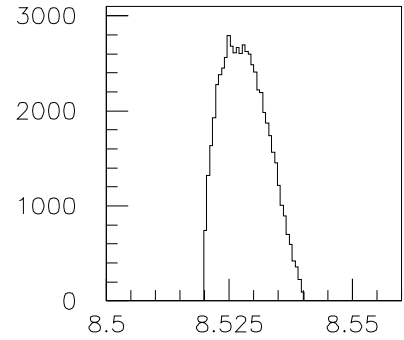
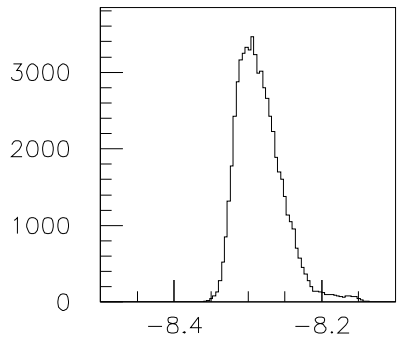
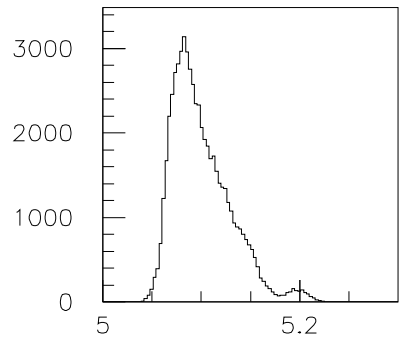
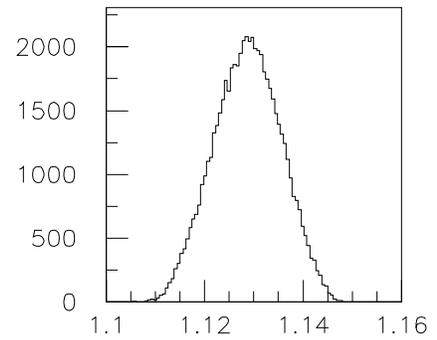
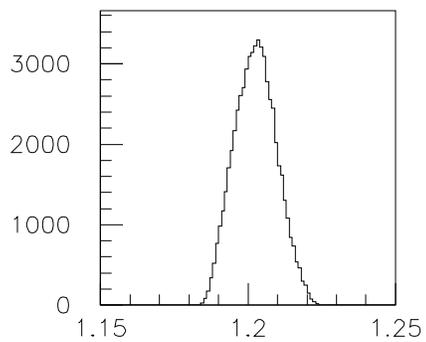
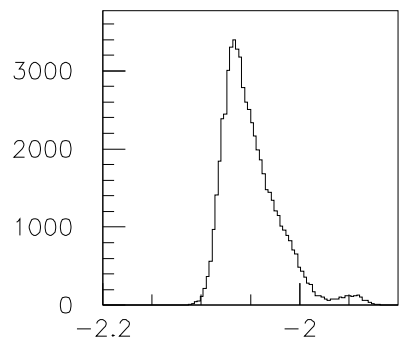
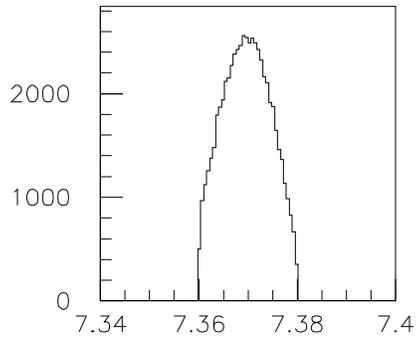
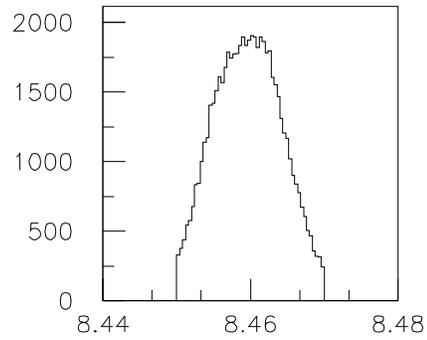
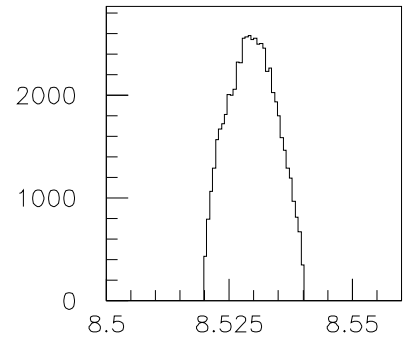
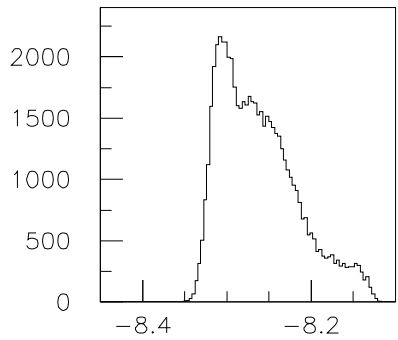
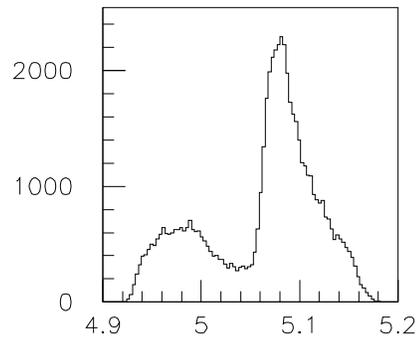
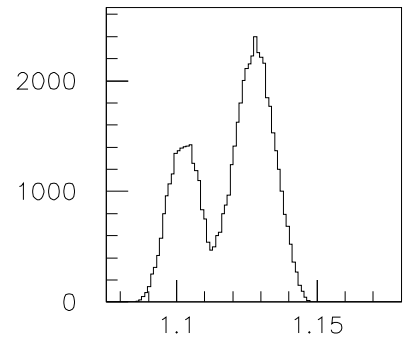
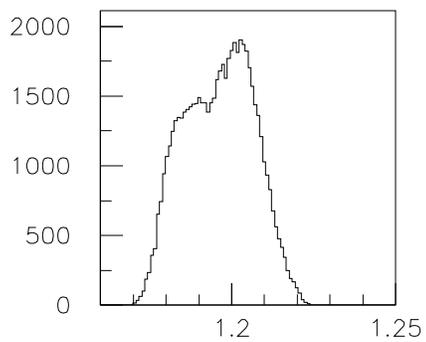
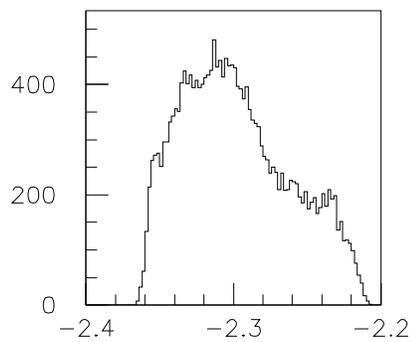
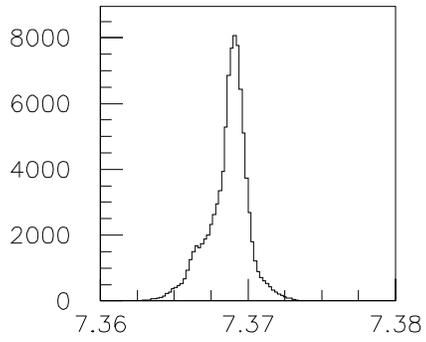
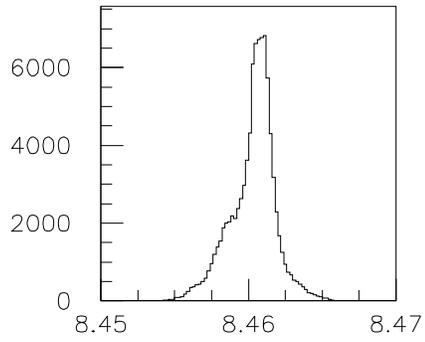
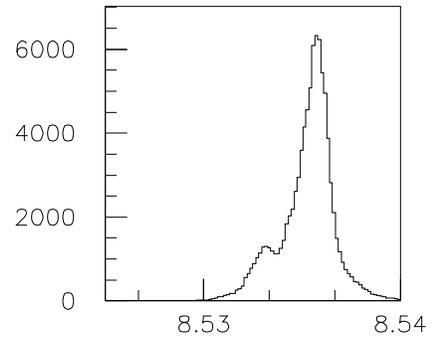
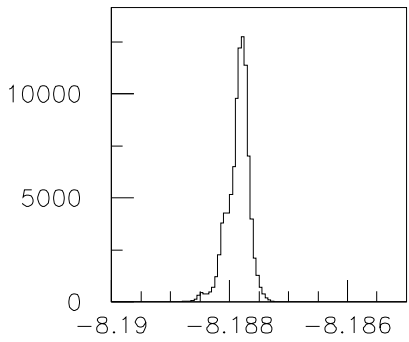
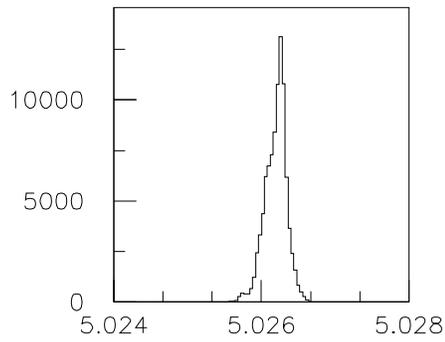
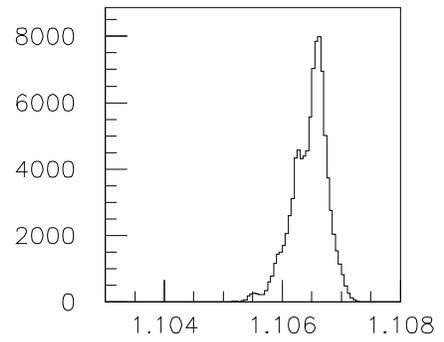
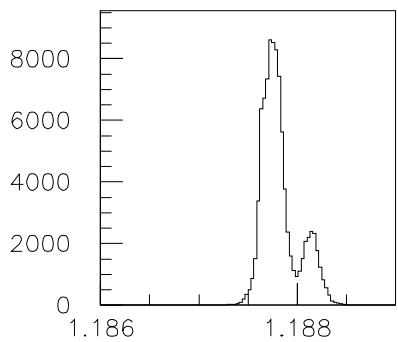
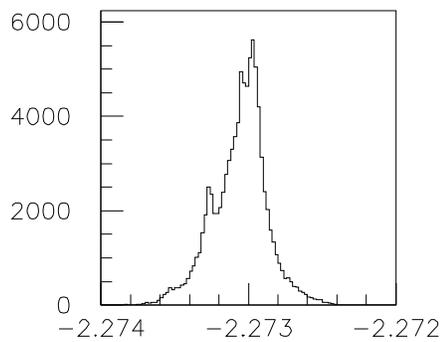
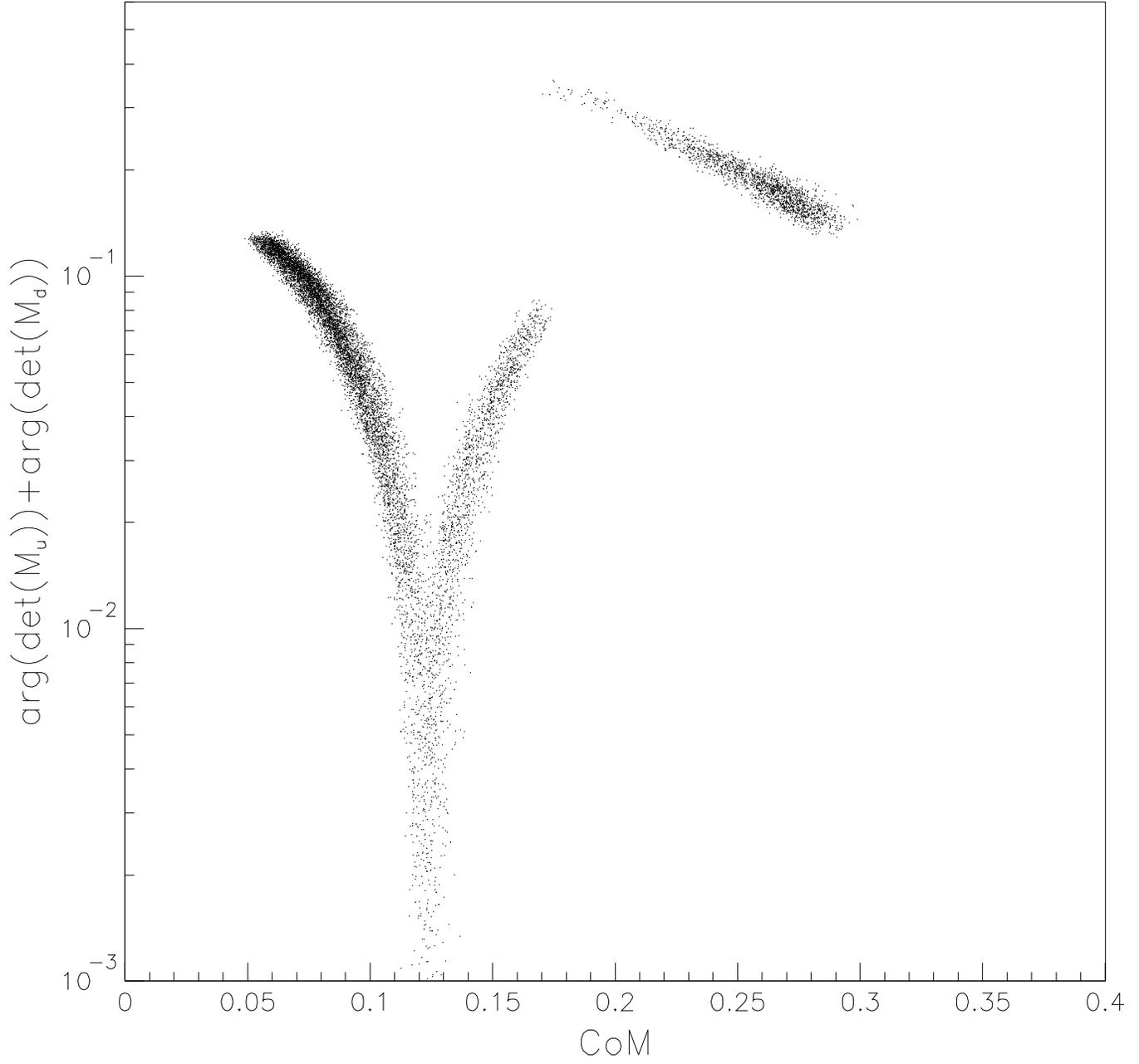
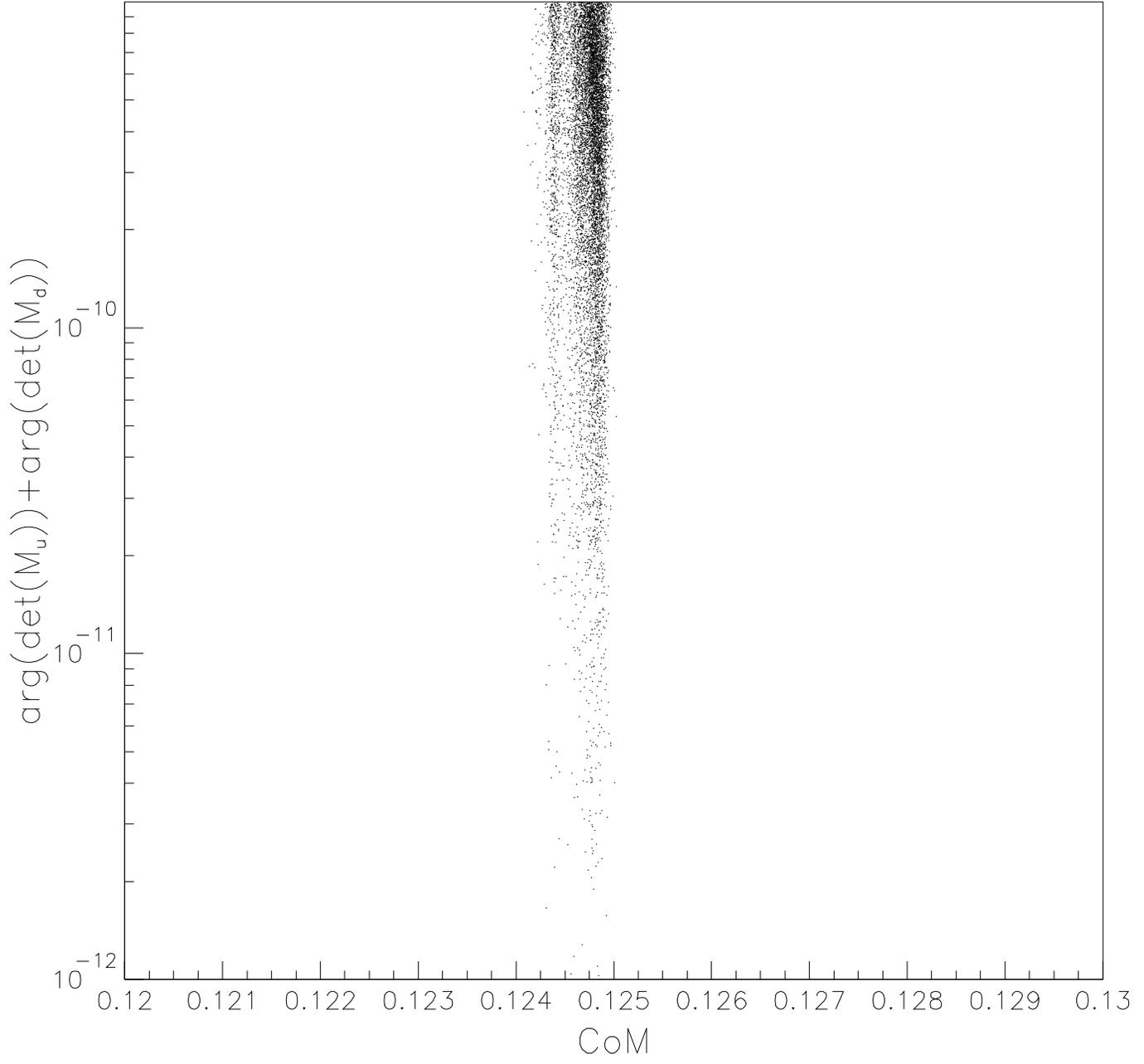
\end{document}